\documentclass{PoS}

\usepackage[utf8]{inputenc}
\usepackage{import}
\usepackage{braket}
\usepackage{multirow}
\usepackage{here}
\usepackage{amsmath}

\usepackage{subfigure}
\title{Follow-up on non-leptonic kaon decays at large $N_c$}

\ShortTitle{Follow-up on non-leptonic Kaon decays at large $N_c$}

 \author{\speaker{Fernando Romero-L\'opez} \\  IFIC, CSIC-Universitat de Val\`encia, \\ 46980 Paterna, Spain \\
     \email{fernando.romero@uv.es}
        }
 \author{Andrea Donini \\  IFIC, CSIC-Universitat de Val\`encia,\\ 46980 Paterna, Spain \\
     \email{donini@ific.uv.es}
        }

        \author{Pilar Hern\'andez \\  IFIC, CSIC-Universitat de Val\`encia,\\ 46980 Paterna, Spain \\
     \email{m.pilar.hernandez@uv.es}
        }
        \author{Carlos Pena \\  Departamento de F\'isica Te\'orica and IFT UAM-CSIC, Universidad Aut\'onoma de Madrid, \\ 28049 Madrid, Spain \\
     \email{carlos.pena@uam.es}
        }

\abstract{We report on the status of our dynamical simulations of a $SU (N_c )$ gauge theory with $N_c=3-6$ and $N_f =4$ fundamental fermions. These ensembles can be used to study the Large $N_c$ scaling of weak matrix elements in the GIM limit $m_c=m_u$, that might shed some light on the origin of the $\Delta I=1/2$ rule.  We present preliminary results for the $K \to \pi$ matrix elements in the $N_c=3$ dynamical simulations, where we observe a significant effect of the quark loops that goes in the direction of enhancing the ratio of $A_0/A_2$ amplitudes. Finally, we present the relevant NLO Chiral Perturbation Theory predictions for the relation between  $K \to \pi $ and $K \to \pi \pi$ amplitudes in the light charm limit. }

\FullConference{The 36th Annual International Symposium on Lattice Field Theory - LATTICE2018\\
		22-28 July, 2018\\
		Michigan State University, East Lansing, Michigan, USA.}

\begin{document}

\section{Motivation}
The Standard Model prediction for non-leptonic kaon decays remains unclear. In particular, there is no satisfactory explanation for the longstanding puzzle of the $\Delta I=1/2$ rule, that is, the large hierarchy between the two isospin weak decay amplitudes $K \to (\pi\pi)_{I=0,2}$ which result in a ratio ${A_0}/{A_2} \sim 22$. In the last few years, there has been important progress in the Lattice QCD computation of these amplitudes. The enhancement has been observed, although large uncertainties are still present \cite{Bai:2015nea,Blum:2015ywa, Ishizuka:2018qbn}.  

In our previous work \cite{Donini:2016lwz, Donini:2017rzi}, we saw that the large $N_c$ limit of the $K \to \pi$ amplitude was the expected one, leading to an isospin ratio of $A_0/A_2=\sqrt{2}$. Moreover, the subleading $1/N_c$ corrections of the two amplitudes were of the natural expected size and fully anticorrelated i.e. pointing towards an enhancement in the ratio, coming from both an enhancement of the $\Delta I=1/2$ amplitude and a suppression of the $\Delta I=3/2$ one. Still, the observed enhancement was not enough to explain the $\Delta I=1/2$ rule. 

One of the main limitations of our study for the $K \to \pi$ amplitudes was the use of the quenched approximation. This approximation is expected to recover the exact Large $N_c$ limit, although it can affect the subleading $1/N_c$ corrections. In addition, our first study was restricted to a heavy kaon mass, so the mass dependence of the matrix elements remained unexplored. 

State-of-the-art simulations involve $N_f=2+1+1$ and physical masses for the quarks. However, dynamical simulations with $N_c>3$ have not been systematically used in the context of zero temperature QCD. Such ensembles can be of general interest not only for weak decays, but also for other observables, such as Low Energy Constants (LECs) of the chiral Lagrangian, scattering parameters, etc. 

In this work, we summarize the status of our simulations with $N_c\geq 3$ and we use the $N_c=3$ ones to calculate the $K \to \pi$ matrix elements. This allows us to quantify the impact of quenching, which turns out to be of 60\% in the ratio of the two $K \to \pi$ amplitudes.

\section{Details of the Simulations}

Our configurations have been generated using the latest version of the HiRep \cite{DelDebbio:2008zf} \footnote{We thank C. Pica and M. Hasen for making the code available.}. We use the Iwasaki gauge action and $O(a)$ improved Wilson fermions on the sea. For $N_c=3$, we use the same gauge parameters as in Ref. \cite{Alexandrou:2018egz}. For the other values of $N_c$, we tune $\beta$ so that the lattice spacing is as close as possible. In addition, the value of $c_{sw}$ is the perturbative one (see Ref. \cite{Aoki:2003sj}) boosted by the plaquette for  $N_c=3$, and it is kept constant for $N_c>3$. This choice is motivated by the fact that
\begin{equation}
c_{sw} = 1 + g^2 c^{(1)}_{sw},
\end{equation}
where $g^2 \sim 1/N_c $ and $c_{sw}^{ {(1)} } \simeq c_{sw}^{\text{tad}} \sim N_c $, as can be seen from Eq. (58) in Ref. \cite{Aoki:2003sj}. Therefore, the previous choice would only have effects involving terms $O(a^2/N_c)$.

In Table \ref{tab:ensembles}, we list the current status of our ensembles. They are complete, apart from ensemble A404, which is still running. Moreover,  for $N_c=5,6$ new ensembles will be available soon.

\begin{table}[h!]
\centering
\begin{tabular}{|c|c|c|c|c|c|c|c|}
\hline
Ensemble & $N_c$              & $L \times T$ & $\beta$                & $m_0$   & \#configs \footnotemark & $aM$ & $t^{\text{imp}}_0/a^2$ \\ \hline \hline
A301     & \multirow{4}{*}{3} & $20\times36$ & \multirow{4}{*}{1.778} & -0.404  & 900  & 0.2191(36) & 3.263(50)      \\ \cline{1-1} \cline{3-3} \cline{5-8} 
A302     &                    & $24\times48$ &                        & -0.406  & 1700 & 0.1831(17)   & 3.491(32)   \\ \cline{1-1} \cline{3-3} \cline{5-8} 
A303     &                    & $24\times48$ &                        & -0.407  & 1400   & 0.1612(24)  & 3.740(39)    \\ \cline{1-1} \cline{3-3} \cline{5-8} 
A304     &                    & $32\times60$ &                        & -0.408  & 900  & 0.1400(18)  & 3.848(34)    \\ \hline \hline
A401     & \multirow{4}{*}{4} & $24\times48$ & \multirow{4}{*}{3.570} & -0.3725 & 900   &  0.2052(37) & 3.494(45)  \\ \cline{1-1} \cline{3-3} \cline{5-8} 
A402     &                    & $24\times48$ &                        & -0.3752 & 1600 & 0.1788(18) & 3.565(26)     \\ \cline{1-1} \cline{3-3} \cline{5-8} 
A403     &                    & $24\times48$ &                        & -0.376  & 1400  & 0.1703(13) & 3.593(29)     \\ \cline{1-1} \cline{3-3} \cline{5-8} 
A404    &                    & $32\times60$ &                        & -0.378  & 200    & 0.1430(9) & 3.705(18)  \\ \hline \hline
A501     & \multirow{1}{*}{5} & $20\times36$ & \multirow{1}{*}{5.969} & -0.3458 & 1000  &0.2130(13) & 3.532(17)     \\  \hline\hline
A601     & 6                  & $20\times36$ & 8.974                  & -0.326  & 700 &  0.2163(8) &3.619(17)       \\ \hline
\end{tabular}
\caption{Summary of the current status of the ensembles. $c_{sw}=1.69$ is kept throughout. The lattice spacing is $a\simeq 0.076\text{ fm}$. $t^{\text{imp}}_0$ is calculated using the tree level improvement of $t_0$ in Ref. \cite{Fodor:2014cpa}.}
\label{tab:ensembles}
\end{table}

\section{Scale Setting at Large $N_c$}

In order to determine the value of the lattice spacing $a$ in physical units, we need to find  a suitable observable. Fermionic observables such as $F_K$ are widely employed, but they depend strongly on $N_f$ and the mass of the quarks. In our case, it seems a complicated task to extrapolate an observable measured with four degenerate flavours to the physical case, where the charm is very heavy. Thus, gluonic observables represent a more sensible choice. In particular, we will use $t_0$ measured with the gradient flow, whose definition at $N_c=3$ is:
\footnotetext{{It refers to configurations with autocorrelation.}}
\begin{equation}
\braket{ t^2 E(t)}\Big|_{t=t_0} = c =0.3.
\end{equation}
The dependence in $N_c$ and $N_f$ of this observable is known from perturbation theory \cite{Luscher:2010iy}:
\begin{equation}
\braket{ t^2 E(t)} = \frac{3}{128 \pi^2} \frac{N_c^2-1}{N_c} \lambda\left(q\right) \left( 1+  \frac{c_1}{4\pi} \lambda(q) + O(\lambda^2) \right),
\label{eq:t0N}
\end{equation}
in terms of the 't Hooft coupling $\lambda(q)$ at the scale $q=1/{\sqrt{8t}}$, and where $c_1=0.36593+0.0075 {N_f}/{N_c}$. From Eq. \eqref{eq:t0N}, we see that the $N_f$ dependence arises at 1 loop and with a small coefficient. Hence, we will generalize $t_0$ to an arbitrary $N_c$ as \footnote{The same idea was used in \cite{Ce:2016awn}.}
\begin{equation}
\braket{ t^2 E(t)}\Big|_{t=t_0} = c(N_c)=\frac{N_c^2-1}{N_c} 0.1125, \label{eq:t0N2}
\end{equation}
which is exact up to a very small $N_f/N_c$ correction ($\sim 0.5 \%$).
We also need the value of $t_0$ in physical units. This is known from lattice simulations for $N_f=2$  \cite{Bruno:2013gha, Sommer:2014mea} and $N_f=3$ \cite{Bruno:2016plf} degenerate quarks and $M_\pi=420\text{ GeV}$.
\begin{equation}
\sqrt{t_0}\Big|^{N_f=2}_{M= 420 \text{ MeV}} = 0.1470(14) \text{ fm,}  \ \ \ \sqrt{t_0}\Big|^{N_f=3}_{ M=420 \text{ MeV}} = 0.1460(19) \text{ fm} 
\end{equation}
 Having that, it is easy to extrapolate to $N_f=4$:
\begin{equation}
\sqrt{t_0}\Big|^{N_f=4}_{ M=420 \text{ MeV}} = 0.1450(39) \text{ fm,}
\end{equation}
and our scale setting condition will involve the dimensionless quantity
\begin{equation}
(M \sqrt{t_0}) \Big|_{M=420 \text{ MeV}} = 0.3091(83). \label{eq:t0Mref}
\end{equation}

Furthermore, Eq. \eqref{eq:t0Mref} requires a chiral extrapolation of $t_0$ to be used. This has been derived in Ref. \cite{Bar:2013ora} using chiral perturbation theory and for the case of degenerate flavours is:
\begin{equation}
t_0  =   t_0^{ch} \left(1 + k \ M^2 \right)  + O(M^4),
\end{equation} 

\begin{figure}[tp]
   \centering
   \subfigure[Ensembles with $N_c=3$ \label{fig:t0Nc3}]%
             {\includegraphics[width=0.475\textwidth,clip]{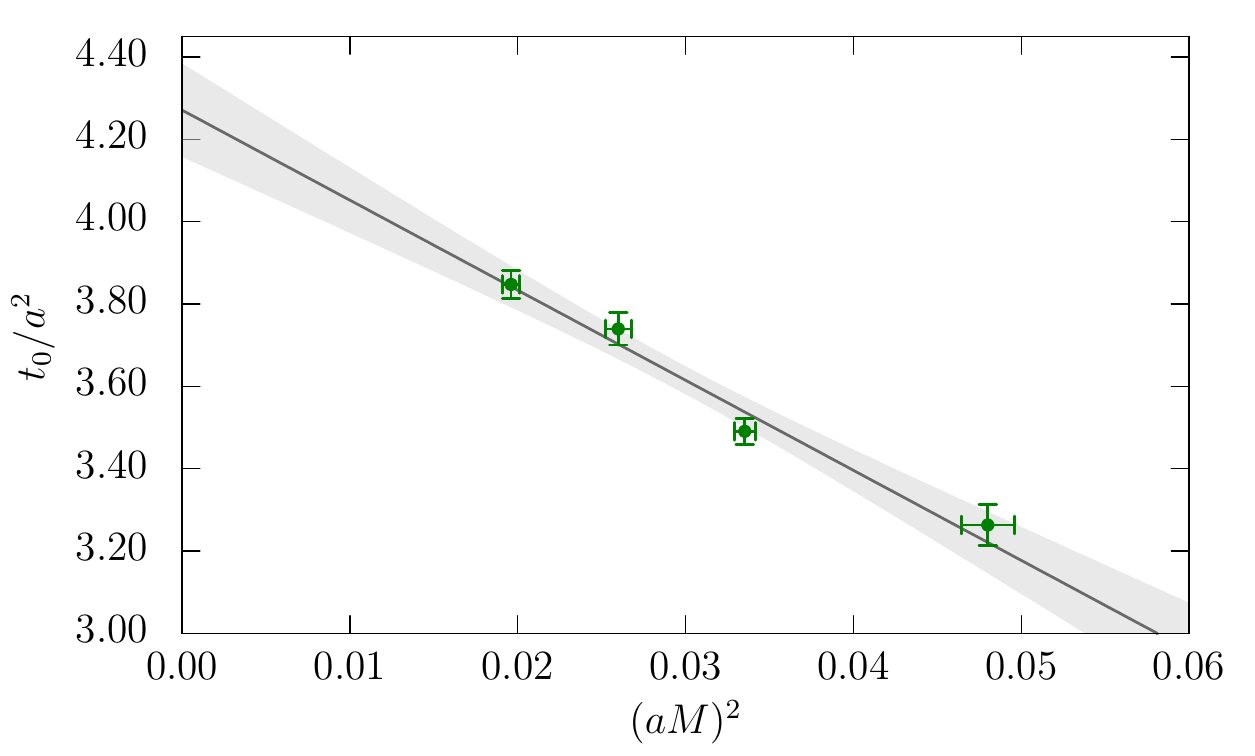}}\hfill
   \subfigure[Ensembles with $N_c=4$  \label{fig:t0Nc4}]%
             {\includegraphics[width=0.475\textwidth,clip]{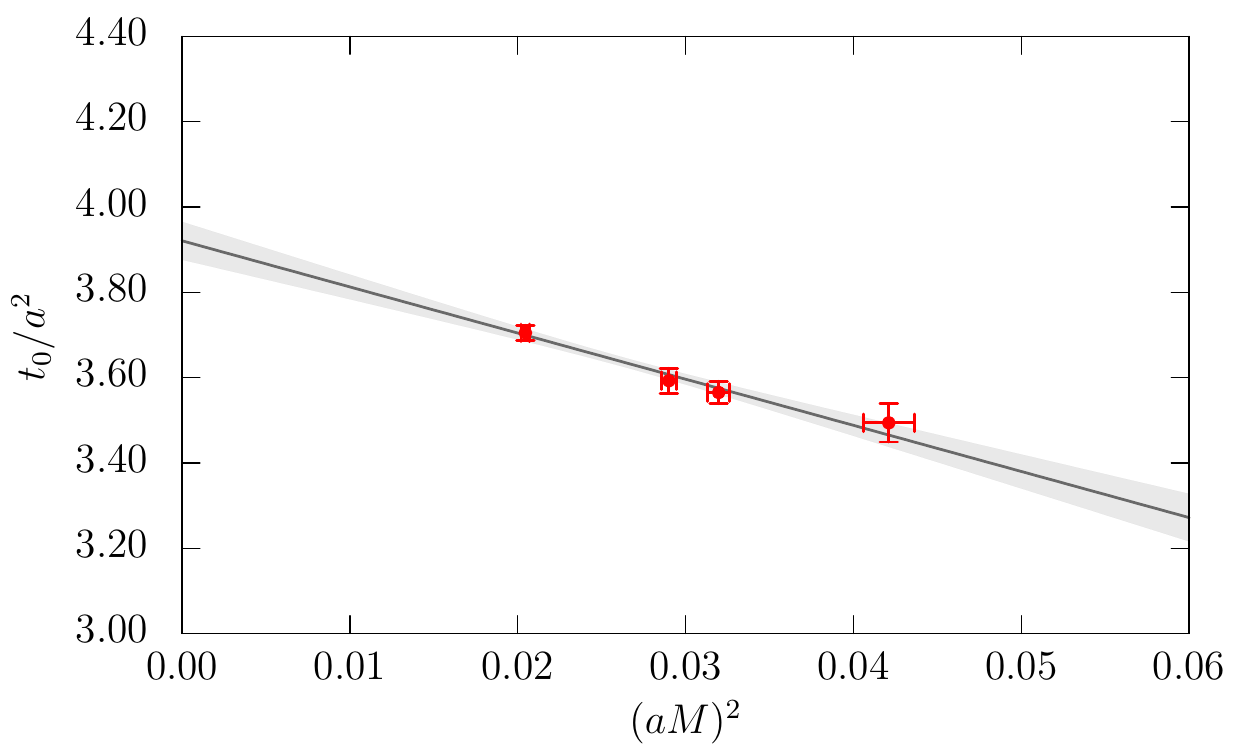}}
   \caption{Chiral extrapolation of $t_0$}
   \label{fig:t0fig}
\end{figure}

In Figs. \ref{fig:t0Nc3} and \ref{fig:t0Nc4}, we show the chiral extrapolation of $t_0$ for $N_c=3,4$. This allows to compute the lattice spacing. It can be seen how increasing $N_c$ yields a flatter mass dependence. For $N_c=5,6$, only one point is available, so we cannot compute the mass dependence yet. Still, we can estimate the lattice spacing by assuming that $t_0$ will not change with the mass, which will be very close to reality for these values of $N_c$.  The results are summarized in Table \ref{tab:a}. Our results show that the lattice spacing is tuned to the percentage level across the values of $N_c$.

\begin{table}[h!]
\centering
\begin{tabular}{|c|c|}
\hline
$N_c$ & $a$ \\ \hline
3     &  $0.0753(4)(20)\text{ fm}$    \\ \hline
4     &  $0.0760(2)(20)\text{ fm}$  \\ \hline 
5     &   $\lesssim 0.0771(2)(21)\text{ fm}$  \\ \hline
6     &    $\lesssim 0.0762(2)(20)\text{ fm}$ \\ \hline
\end{tabular}
\caption{Results for the scale setting. The first error is statistical and the second comes from the uncertainty in $t_0$ in physical units. For $N_c=5,6$ the lattice spacing is just an upper bound, since we have used $t_0$ measured at a higher mass $M_1$, such that $t_0(M_1)>t_0(M=420\text{ GeV})$. The mass dependence of $t_0$ is suppressed in the Large $N_c$ limit, so this effect is expected to be of the order 1\%.  }
\label{tab:a}
\end{table}

\section{Kaon Matrix Elements}

One reason to generate the ensembles of the previous section was to study the impact of the quenched approximation in the $K \to \pi$ matrix element, which can be connected to $K \to \pi \pi$ via chiral perturbation theory. We will use the mixed-action approach (see Ref. \cite{Herdoiza:2017bcc}), with twisted mass fermions at maximal twist in the valence sector. In this approach, the pseudoscalar mass in the valence sector matches the pseudoscalar mass in the sea with Wilson fermions. 

 It is known that quenching has a small effect in other quantities, such as $F_\pi$. Nevertheless, its systematics are unknown and different for various observables. Therefore, for $K \to \pi$ they can be relevant and they can alter the subleading $1/N_c$ corrections. At this stage of the project, we can only present results for $N_c=3$. For the complete formalism, we refer to our previous work \cite{Donini:2016lwz}, in particular, for the calculation of the Wilson coefficients and renormalization constants. Here we will just summarize the most important part. 

We start with the $N_f=4$ effective weak Hamiltonian
\begin{gather}
\label{eq:heffs1}
H_{\rm w}^{\Delta S=1} = \int d^4x~\frac{g_{\rm w}^2}{4M_W^2}V_{us}^*V_{ud}\sum_{\sigma=\pm} k^\sigma(\mu) \, \bar{Q}^\sigma(x,\mu)\,,
\end{gather}
where $k^\sigma(\mu) = k^\sigma(M_W) \, U^\sigma(\mu,M_W) $ are the Wilson coefficients at the scale $\mu$ and
\begin{gather}
\begin{split}
{\bar Q}^\pm(x,\mu) = Z_Q^\pm(\mu) \, \big(& J_\mu^{su}(x)J_\mu^{ud}(x) \pm J_\mu^{sd}(x)J_\mu^{uu}(x)
~-~[u\leftrightarrow c]\big)\, , \label{eq:ops}
\end{split}
\end{gather}
are the effective operators with the renormalization constants, $Z_Q^\pm(\mu)$   . 
With that, we calculate the following ratios on the lattice
\begin{eqnarray}
\bar{R}^\pm \equiv \frac{\langle\pi|\bar{Q}^\pm|K\rangle}{f_Kf_\pi m_Km_\pi}
= Z_R^\pm(\mu) R^\pm_{bare} \,,
\label{eq:ratio}
\end{eqnarray}
and the full $K \to \pi$ amplitudes\
\begin{equation}
A^\pm = k^\pm(\mu) \bar{R}^\pm = k^\pm(M_W) \, U^\pm(\mu,M_W) Z_R^\pm(\mu) R^\pm_{bare}. \label{eq:apm}
\end{equation}
The numerical values of the coefficients in Eq. \eqref{eq:apm} are given in Table \ref{tab:coeffs}. With them, we calculate the ratio of the amplitudes that we can compare to  our previous quenched result:
\begin{equation}
\frac{A^-}{A^+}\Big|_{N_c=3,N_f=4} = 4.1(3), \ \ \ \ \frac{A^-}{A^+}\Big|_{N_c=3,N_f=0} = 2.4(1).
\end{equation}
It can be seen that the effect is of around $60\%$.

The last step is relating $A^\pm$ to $A_I = \braket{K | \left(H_{ w}^{\Delta S=1}\right)_I | \pi \pi}$ with $I=0,2$. This had been done previously to leading order in Chiral Perturbation Theory (see Ref. \cite{Giusti:2006mh}). We present here the NLO result:
\begin{align*}
{ \text{Re } \frac{A_0}{A_2} \Big \rvert_{M_\pi,M_D \to0, M^{\text{phys}}_K}}   = {\frac{1}{2\sqrt{2}} \left( 1 + 3\frac{A^-}{A^+} \right)  } {- \frac{17}{6\sqrt{2}}  \left( 1+ \frac{1}{17}\frac{A^-}{A^+} \right) \frac{M_K^2}{(4 \sqrt{2} \pi F)^2} \log \frac{M_K^2}{\mu_{eff}^2}}  
\end{align*}
where $\mu_{eff}$ is an unknown scale that contains information of the NLO LECs of the effective Chiral Hamiltonian. 

The result for the ratio, with a reasonable guess of $\mu_{eff}$ between  $M^{phys}_\rho$ and $ 2 \text{ GeV}$, is
\begin{equation}
{ \text{Re } \frac{A_0}{A_2} \Big \rvert_{M_\pi,M_D \to0, M^{\text{phys}}_K}}   = 5.3(4)_{stat}(3)_{\mu_{eff}}.
\end{equation}

\begin{table}[h!]
\centering
\begin{tabular}{|c|c|c|c|c|c|}
\hline
Operator        & $k_1^\pm (M_W)$ & $U^\pm(M_W,\mu)$ & $Z_R^\pm(\mu)$ &$k_1^\pm \ U^\pm \  Z_R^\pm$& $A^\pm_{chiral}$ \\ \hline
$Q^+$ &    1.042             &       0.819           &   1.035   & 0.883   &  0.448(36)           \\ \hline
$Q^-$ &     0.917            &     1.464             &   0.931   &   1.250   & 1.852(39)          \\ \hline
\end{tabular}
\caption{Numerical coefficients needed for the full $K \to \pi$ amplitudes. The last column is the chiral extrapolation of the data points in Fig. \ref{fig:ratiomass}. }
\label{tab:coeffs}
\end{table}

\begin{figure}[tph]
   \centering
   \subfigure[Comparison of $A^\pm$ with the results in the quenched approximation from Refs. \cite{Donini:2016lwz} (black) and  \cite{Donini:2017rzi} (red). \label{fig:ApAm}]%
             {\includegraphics[width=0.475\textwidth,clip]{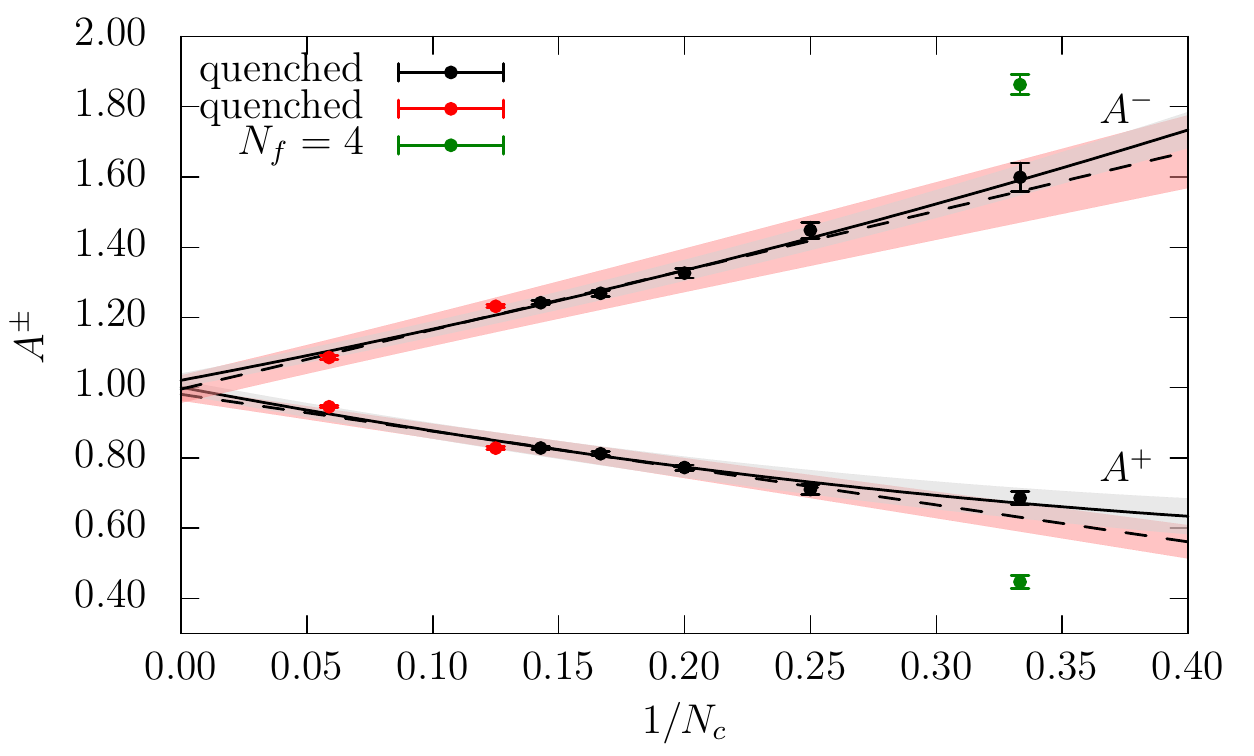}}\hfill
   \subfigure[Lattice results of the $A^\pm$ matrix elements as a function of the mass.  We include the chiral extrapolation. \label{fig:ratiomass}]%
             {\includegraphics[width=0.475\textwidth,clip]{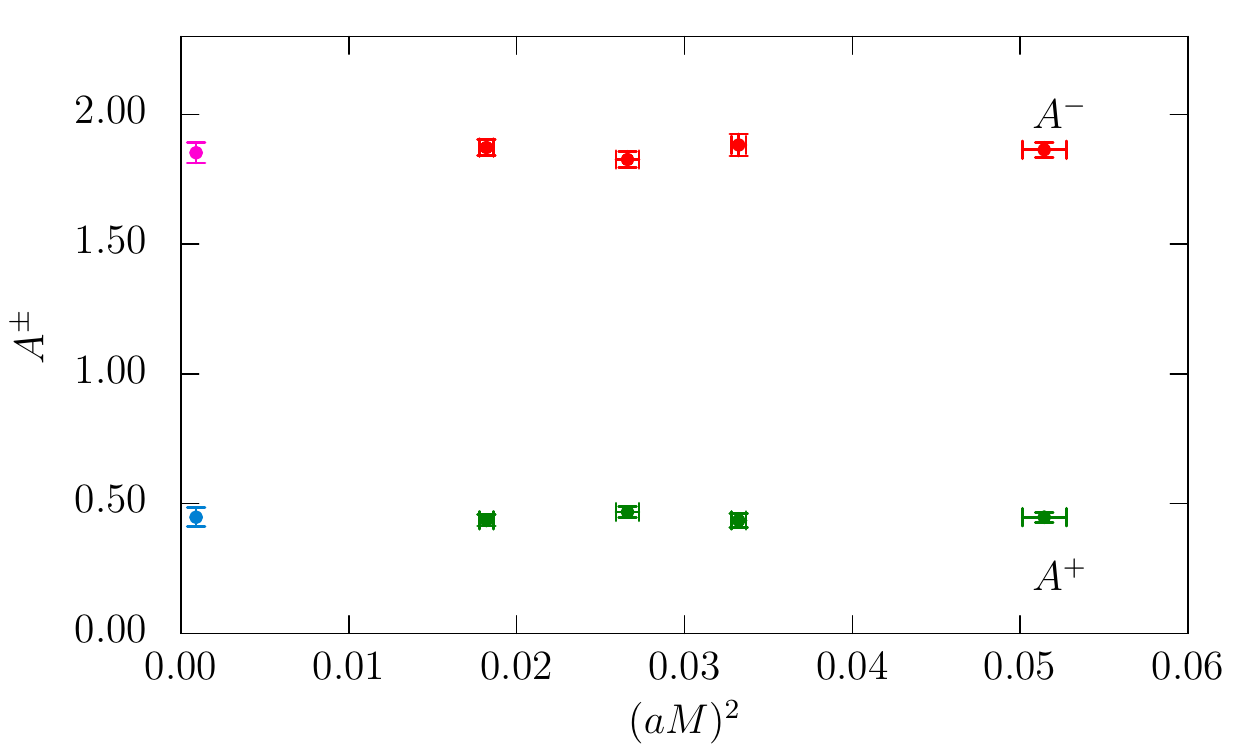}}
   \subfigure[Half-difference of the amplitudes. This is expected to be $O(1/N_c)$ and the effects of quenching are relevant. \label{fig:AmmAp}]%
             {\includegraphics[width=0.475\textwidth,clip]{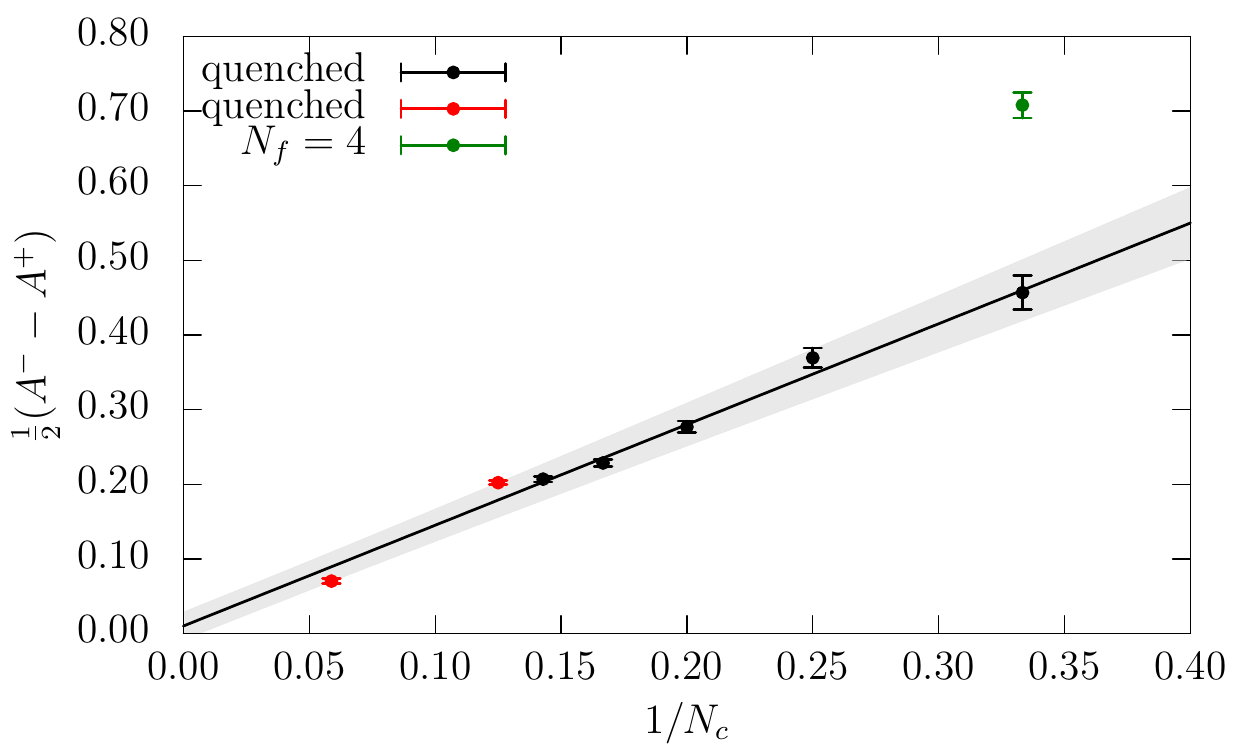}}\hfill
   \subfigure[Half-sum of the amplitudes. This is expected to be $O(1)$. Quenched results agree with $N_f=4$.  \label{fig:AmpAp}]%
             {\includegraphics[width=0.475\textwidth,clip]{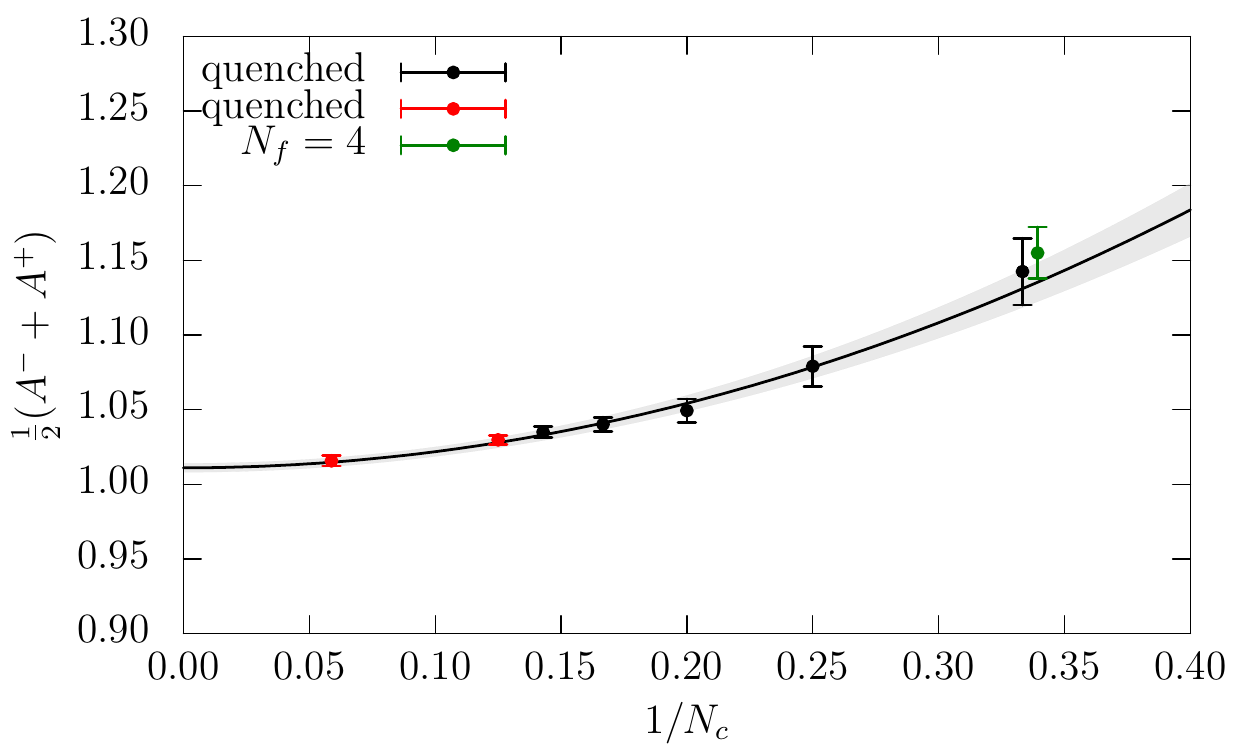}}
   \caption{Numerical results for $K \to \pi$}
   \label{fig:nume}
\end{figure}

\section{Conclusions and Outlook}

In this work we have presented our dynamical simulations of a $SU(N_c)$ gauge theory with $N_c>3$ and $N_f=4$ fundamental fermions. The first motivation for such ensembles was the study of the $N_c$ scaling of the weak decays of kaons, which can have implications for the understanding of the $\Delta I =1/2$ rule. Whereas we recovered the correct large $N_c$ limit in the quenched approximation, we have seen that quenching had an important effect at $N_c=3$ and the results from dynamical simulations have a further enhancement. The effect of quark loops seems to affect sizeably only the leading $1/N_c$ corrections (see Figs. \ref{fig:AmmAp} and \ref{fig:AmpAp}). Our current result is still far from the experimental value, $\text{Re }( {A_0}/{A_2}) \simeq 22$, but some limitations are still present, among them the usage of chiral perturbation theory to relate $A^\pm$ with $A_2,A_0$. In future publications, we will include results with $N_c>3$, and in long term perspective, we intend to calculate the full amplitude, $K \to \pi\pi$.

In addition, these ensembles can help understanding the systematics of the large $N_c$ limit, which is often invoked in phenomenological approaches. In particular, we refer to quantities such as scattering parameters or low energy constants of the chiral Lagrangian. We plan to study these in future works. Finally, we have intention to make our ensembles public.

\section{Acknowledments}
We acknowledge the support of the European Project InvisiblesPlus H2020-MSCA-RISE-2015 and Elusives H2020-MSCA-ITN-2015//674896-ELUSIVES020-MSCA-ITN-2015. FRL, AD and PH have also received funding through the MINECO project FPA2017-85985-P. In addition, the work of FRL has received funding from the European Unions Horizon 2020 research and innovation program under the Marie Sklodowska-Curie grant agreement No. 713673 and from La Caixa foundation. CP acknowledges support through the Spanish MINECO project FPA2015-68541-P and the Centro de Excelencia Severo Ochoa Programme SEV2016-0597. These simulations have been possible with the resources granted by Finis Terrae II (CESGA), Tirant (UV) and Lluis Vives (UV).

\end{document}